\documentclass[runningheads]{llncs}
\usepackage[T1]{fontenc}
\usepackage{algorithm}
\usepackage{algorithmic}
\usepackage{amsmath}
\usepackage{amssymb}
\usepackage{tabularx} 
\usepackage{graphicx} 
\usepackage{hyperref}
\newtheorem{assumption}{Assumption}
\newcommand{\INPUT}{\item[\textbf{Input:}]}
\newcommand{\OUTPUT}{\item[\textbf{Output:}]}

\usepackage{xcolor}
\begin{document}
\title{Time Series Treatment Effects Analysis with Always-Missing Controls}
%
%
%
\author{Juan Shu\inst{1*} \and
Qiyu Han\inst{1*} \and George Chen\inst{1} \and Xihao Cao\inst{1} \and Kangming Luo\inst{1} \and Dan Pallotta \inst{1} \and Shivam Agrawal \inst{1} \and Yuping Lu \inst{1} \and Xiaoyu Zhang \inst{1} \and Jawad Mansoor \inst{1} \and 
Jyoti Anand\inst{1}}
\authorrunning{J.Shu et al.}
%
\institute{Walmart Global Tech \\
\email{juan.shu,qiyu.han,george.chen0,xihao.cao,kangming.luo,daniel.pallotta,\\
shivam.agrawal, yuping.lu, xiaoyu.zhang1, jawad.mansoor,jyoti.anand@walmart.com}\\}
\maketitle                 
\begin{abstract}
Estimating treatment effects in time series data presents a significant challenge, especially when the control group is always unobservable. For example, in analyzing the effects of Christmas on retail sales, we lack direct observation of what would have occurred in late December without the Christmas’s impact. To address this, we try to recover the control group in the event period while accounting for confounders and temporal dependencies. Experimental results on the M5 Walmart retail sales data demonstrate robust estimation of the potential outcome of the control group as well as accurate predicted holiday effect. Furthermore, we provided theoretical guarantees for the estimated treatment effect, proving its consistency and asymptotic normality. The proposed methodology is applicable not only to this always-missing control scenario but also in other conventional time series causal inference settings.

\end{abstract}

\section{Introduction}
\footnote{* Equal contribution.}
Time series treatment effect estimation holds importance in a variety of practical applications: policy changes, marketing campaigns, and seasonal event monitoring (e.g. holidays). A significant challenge arises when the control group is inherently always unobservable. For example, we cannot observe consumer behavior during Independence Day in absence of the holiday itself. Practitioners must rely on data from non-holiday periods to learn about the control group's properties, making it essential to develop methodologies that can accurately estimate treatment effects under these constraints.

Our work addresses a critical challenge in the context of causal inference for time series, where the control group is always unobservable during the treatment period. Existing studies in time series causal inference often rely on assumptions such as the availability of pre-treatment and post-treatment data for both treated and control groups, or they focus on methods like difference-in-differences, which require observable controls throughout the study period \cite{abadie2010synthetic,imbens2015causal,athey2021matrix,brodersen2015causalimpact}. These approaches fall short in scenarios where the control group cannot be observed, such as during holiday seasons. To the best of our knowledge, our framework is the first work to bridge this gap by leveraging predictive modeling from non-treatment periods and integrating it with causal inference principles, providing a robust solution tailored to scenarios where the control group is always missing during treatment.

The approach is inspired by anomaly detection using generative models, where primary patterns are effectively captured, but anomalies are challenging to reconstruct. Rare events, particularly in daily or higher-frequency time series, can be treated as anomalies. Therefore, we use deep learning models to fit the entire time series and attempt to recover it with insample forecasting. The fitted values in regions corresponding to rare events are treated as synthetic controls, simulating a scenario without the rare events. This approach assumes that the sparse signals of rare events will not be reconstructed by the model and left over in the residuals. To adddress, we have an adaptive loss adjusting the weights of rare events in each iteration when we train the models. 

Several studies have explored related problems in dynamic panel data models and treatment effect estimation. \cite{arellano1991some} and \cite{blundell1998initial} developed instrumental variable and GMM estimators for dynamic panels, particularly effective when both cross-sectional ($N$) and time ($T$) dimensions are large. \cite{hsiao2002panel} provided comprehensive methods for analyzing panel data, focusing on cases with short time dimensions. In the context of causal inference, \cite{angrist2008mostly} and \cite{imbens2015causal} extensively discuss econometric techniques for estimating causal effects, though these methods often rely on observable controls during the treatment period. Our contribution extends this literature by addressing the challenge of estimating treatment effects when the control group is unobservable during the treatment period, leveraging cross-sectional information from non-treatment periods. One key advantage of our approach is its flexibility, as it does not require prior knowledge of the rare event window size, allowing for broader applicability. The contributions of the paper can be summarized as:
\begin{enumerate}
    \item We introduce an innovative estimation method specifically designed for time series causal inference where the control group is always unobservable during the treatment period—a situation not adequately handled by existing time series and causal inference methodologies.
    
    \item We provide rigorous theoretical guarantees by proving the asymptotic properties of our causal estimator.

    \item Empirical analyses on actual retail sales data from Walmart validate our method, demonstrating the accuracy and robustness of our estimator. 
\end{enumerate}

\section{Methodology}
\subsection{Preliminaries}
We aim to estimate the causal effect of rare events in time series data. A rare event, denoted by $\mathcal{R}$, is characterized by a window of size $|T_\mathcal{R}| = d$, where $d$ is negligible relative to the overall observation period. In this study, the rare events of interest recur annually within the same time window. Holidays serve as a typical example of such events in our context. Following the notation convention in the potential outcomes framework (\cite{rubin2005causal}), we denote $Y_t(1)$ for any $t \in T_\mathcal{R}$ as the observed time series during the event period. Conversely, $Y_t(0)$ for any $t \notin T_\mathcal{R}$ represents the time series observations during the non-event period.

\subsection{Estimating synthetic control}
 Statistical models and deep learning models are primarily designed to capture the main signal from input data, often leaving anomalies in the residuals. This characteristic is essential for many anomaly detection techniques. For example, confidence interval-based methods fit models with uncertainty quantification, flagging anomalies when new data points fall outside the predetermined range. Similarly, autoencoders assess reconstruction errors, identifying anomalies as data points that are difficult to accurately reconstruct. These anomaly detection techniques are not only useful for identifying outliers but also offer insights into constructing synthetic controls which serve as a baseline to compare with the impact of interventions or events. By analyzing the residuals or reconstruction errors, we can estimate what the expected ``normal'' behavior should be under specific conditions. This methodology can then be extended to create synthetic controls, which serve as a baseline to compare and assess the impact of interventions or events. 

Formally, we trained a deep learning model with an adaptive loss function to adjust the loss landscape in rare even regions. We then performed in-sample forecasts to extract the synthetic control. The estimated synthetic control can be expressed as:
\begin{equation}
    \{\hat{Y}_{i,t}(0)\}_{t\in T_\mathcal{R}}=\hat{f}_\theta(X_{tr}),
\end{equation}
where $f_\theta(\cdot)$ is the trained deep learning model parametrized by $\theta$, with adaptive loss function defined in \eqref{eq: loss}. $X_{tr}$ is the training data. $\{\hat{Y}_{i,t}(0)\}_{t\in T_\mathcal{R}}$ s the estimated synthetic control during rare event region $\mathcal{R}$.

\begin{algorithm}
\caption{Rare event impact extraction}
\label{alg:weighted_loss}
\begin{algorithmic}[1]  
\INPUT $\{Y_t\}_{t=0}^{T}$, $s$, $T_\mathcal{R}$, $H$, $M$.

\OUTPUT event impact estimate $\hat{\delta}$

\STATE Create training data $Y_i^{tr} = \{y_o\}_{o=i\times s}^{i\times s+M}$ and $Y_i^{label} = \{Y_j\}_{j=i\times s+M}^{i\times s+M + H}$, $i \in \{0,1,\dots, T-H\}$.
\STATE Initialize model $f_\theta$
\FOR{$epoch = 1$ to $N$}
    \FOR{batch $(Y_i^{tr}, Y_i^{label})$ in \textbf{BATCH}}
        \STATE $\hat{Y}_i^{label} = f_\theta(Y_i^{tr})$.
        \STATE Compute weighted loss $\mathcal{L}_i$.
        \STATE Update model $\theta \gets \theta - \alpha \nabla_\theta \mathcal{L}_i$.
    \ENDFOR
\ENDFOR
\STATE Extract rare event effect estimate $\hat{\delta} = g(\{\hat{Y}_i\}_{i=H+1}^{T-H}, \{Y_i\}_{i=1}^T, T_\mathcal{R})$.
\end{algorithmic}
\end{algorithm}

\subsection{Temporal Adaptive loss}
 A rare event usually has larger loss compared with non-rare events region. Therefore, in order to ensure a good landscape of loss function, and capture the synthetic control more accurately, we train the model with a temporal adaptive loss function defined in \eqref{eq: loss}. 

\begin{equation}
\label{eq: loss}
    \mathcal{L} = \frac{1}{N}\sum_{i=1}^{N}\left(\boldsymbol{w}_{1,i}\sum_{t \in T^i_r}\eta(Y_{i,t}, \hat{Y}_{i,t}) + \boldsymbol{w}_{2,i}\sum_{t \in T_{-r}}\eta(Y_{i,t}, \hat{Y}_{i,t}) \right)
\end{equation}
 where $\boldsymbol{w}_{1,i}$ and $\boldsymbol{w}_{2,i}$ are the weight vectors applied to the rare events window $T_r$ and non-rare events window $T_{-r}$. As time series training data is created with a rolling manner, where different training sets might cover different subset of rare event regions $|T^i_r|\leq|T_r|$, therefore, different training samples have different weights. In addition,  $\eta(\cdot, \cdot)$ can be any distance function such as MAE loss. To further formalize the process, the training procedure is outlined in Algorithm \ref{alg:weighted_loss}, where the model is trained based on the temporal adaptive loss function.

\section{Theoretical Properties}
In this section, we provide a detailed illustration of the methodology using a time series that follows an AR(1) model during the non-event period. It is important to note that the AR(1) model is used solely as a simplified example for theoretical development. However, the theoretical guarantees remain valid when extended to an ARIMA$(p,\Delta,q)$ model with fixed values of $p$, $\Delta$, and $q$.

\subsection{Model Specification for Treatment Effect}

Consider $N$ independent time series $\{Y_{i,t}\}_{t=1}^T$, where $i = 1, \dots, N$ indexes the each series and $t = 1, \dots, T$ indexes time periods. The model can be written as
\begin{equation}
\label{eq:ar1}
    Y_{i,t} = \phi Y_{i,t-1} + \varepsilon_{i,t},
\end{equation}
where $t = 1, \dots, T_0$, for $T_0$ represents the last time step before the holiday, and $\varepsilon_{i,t}$ denote random noise. During the holiday/treament period ($t = T_0 + 1, \dots, T_0 + d$), the observation $Y_{i,t}$ is given by 
\begin{equation}
\label{eq: model}
Y_{i,t} = Y_{i,t}(1) = Y_{i,t}(0) + \delta_t,
\end{equation}
where $Y_{i,t}(0)$ for $t\in T_{r}$ is the counterfactual outcome without treatment, which follows the same AR(1) process as in the pre-treatment period specified in \eqref{eq:ar1}. Finally, the $\delta_t$ denotes the treatment effect at time $t$, forming the treatment effect vector $\delta = [\delta_{T_0 + 1}, \dots, \delta_{T_0 + d}]^\top$.

To estimate the treatment effects, we first need an estimate of the AR(1) parameter $\phi$ since this parameter will affect our estimation for the control group (counterfactual outcomes) during the rare events window. This estimation is achieved using OLS based on the pre-treatment data. We use the observations $\{Y_{i,t}\}_{t=1}^{T_0}$ for all $i = 1, \dots, N$ to fit an AR(1) model by maximizing the likelihood function, yielding the OLS estimator for $\phi$, which is expressed as
\begin{equation*}
    \hat{\phi} = \frac{\sum_{i=1}^N \sum_{t=2}^{T_0} Y_{i,t-1} Y_{i,t}}{\sum_{i=1}^N \sum_{t=2}^{T_0} Y_{i,t-1}^2}.
\end{equation*}

\subsection{Forecasting Counterfactual Outcomes and Estimating the Treatment Effects}

With $\hat{\phi}$ estimated, we forecast the counterfactual outcomes $\hat{Y}_{i,t}(0)$ for the treatment period, representing what the outcomes would have been in the absence of treatment. At time $t = T_0$, we have $\hat{Y}_{i,T_0}(0) = Y_{i,T_0}$, then for $t = T_0 + 1, \dots, T_0 + d$ we estimate the center factual by
\begin{equation*}
    \hat{Y}_{i,t}(0) = \hat{\phi} \hat{Y}_{i,t-1}(0).
\end{equation*}
Intuitively, using the AR(1) model estimated from the pre-treatment data, we project the expected trajectory of the counterfactual control group. The recursive aspect arises from the fact that only data from outside the holiday period is employed to estimate the parameters. We then use the estimated $\hat{\phi}$ to iteratively forecast the counterfactual values within the holiday period.  

With the estimation of $\phi$ along with the estimation of the counterfactual outcome, the treatment effect at each time $t \in T_r$ is estimated by comparing the observed outcomes to the forecasted counterfactuals. Specifically, we have 
For each $t = T_0 + 1, \dots, T_0 + d$:
\begin{equation}
\hat{\delta}_t = \bar{Y}_t - \hat{\bar{Y}}_t(0) = \frac{1}{N} \sum_{i=1}^N \left( Y_{i,t} - \hat{Y}_{i,t}(0) \right),
\end{equation}
where $\bar{Y}_t:= (1/N) \sum_{i=1}^N Y_{i,t}$ is the sample mean of the observed outcomes, and $\hat{\bar{Y}}_t(0) := (1/N) \sum_{i=1}^N \hat{Y}_{i,t}(0)$ is the sample mean of the estimated counterfactual outcomes.
Therefore, $\hat{\delta}_t$ represents the estimated average effect of the treatment at time $t$ over $N$ independent series.

\subsection{Theoretical Guarantees}

In this section, we provide the theoretical guarantees for the estimation of causal effects when the AR(1) model is specified. In particular, we establish the asymptotic normality of $\hat{\delta}$, and assume the following. 

\begin{assumption}
\label{assum:station}
    \textbf{Stationarity of the AR(1) Process}: The autoregressive parameter satisfies $|\phi| < 1$, ensuring that the process is stationary and has a finite variance.
\end{assumption}

\begin{assumption}
\label{assum:error}
    \textbf{Error condition}: The error term $\varepsilon_{i,t}$ is an \emph{i.i.d.} Gaussian random variable with mean $0$ and variance $\sigma^2$ for all $i$ and $t$.
\end{assumption}

 First, stationarity guarantees as the sample size increases, the sample mean and variance converge to the true mean and variance \cite{brockwell1991time,hamilton1994time}. In terms of error condition, it guarantees that the variability in the data is well-behaved, facilitating the use of asymptotic results \cite{box2015time}. We now present the main theoretical result regarding the estimator $\hat{\delta}$.

\begin{theorem}
    \label{thm:asymptotic_normality}
    \textit{As $N \to \infty$, the estimator $\hat{\delta} = [\hat{\delta}_{T_0 + 1}, \dots, \hat{\delta}_{T_0 + d}]^\top$ satisfies}
    \begin{equation}
    \sqrt{N} (\hat{\delta} - \delta) \xrightarrow{d} \mathcal{N} \left( \mathbf{0}, \Sigma \right),
    \end{equation}
    \textit{where $\delta = [\delta_{T_0 + 1}, \dots, \delta_{T_0 + d}]^\top$ is the true treatment effect vector, and $\Sigma$ is a $d \times d$ diagonal covariance matrix with elements}
    \begin{equation}
    \Sigma_{i,j} = \frac{\sigma^2}{1 - \phi^2} \delta_{i,j}, ~~\text{for } {i,j} \in [d]\times [d], 
    \end{equation}
    \textit{and $\delta_{i,j} = I\{i=j\}$ for $I(\cdot)$ as an indicator function.}
\end{theorem}
Theorem \ref{thm:asymptotic_normality} indicates that the estimator $\hat{\delta}$ is asymptotically unbiased and is normally distributed around the true treatment effect vector $\delta$. Note that the limiting covariance matrix $\Sigma$ is diagonal due to the fact that, asymptotically, the different time points in the treatment period are uncorrelated if the stationarity condition is met. The variance of each component $\hat{\delta}_t$ is $\sigma^2/(1 - \phi^2)$, reflecting both the variability in error terms and the persistence in the AR(1) process.

\section{Experiment}
We used the Walmart retail sales dataset to evaluate the effectiveness of our method in estimating holiday treatment effect during several holidays. We try to answer the following questions with the experiments:
\begin{itemize}

    \item Does our construction of the synthetic control via in-sample forecast perform better than the out-of-sample forecast? \vspace{2mm}
    
    \item Does the estimated synthetic control have the consistency across different years, which empirically supports our established statistical property.  \vspace{2mm}
    
    \item Can estimated synthetic control be generalized well empirically? \vspace{2mm}

    \item Can the method be applicable to different holidays, while preserving decent robustness and accuracy?
\end{itemize}

\subsection{Setup}
The Walmart M5 data are accessible at \hyperlink{https://www.unic.ac.cy/iff/research/forecasting/m-competitions/m5/}{UNIC website}. Original data, provided at the item level, were aggregated to the department level. Additionally, we combined data from three geographical regions: California, Texas, and Wisconsin. This resulted in six daily time series from January 2011 to April 2016, each representing a specific department. Our backbone model is NHITS \cite{challu2023nhits}. The forecast horizon is 30 days and the lookback window is 90 days. We used the Python package Neuroforecast (1.2) and Statsmodels (0.14) in the experiments.

To the best of our knowledge, there are no methods that designed to handle the causal inference problem in always-missing control scenario in time series analysis. However, we did several ablation studies to further demonstrate the effectiveness of our methods.
\begin{itemize}
    \item Direct forecast (DF) : Instead of including the holiday period in the training, we use the data till the start of the holiday as training and forecast the synthetic control within the holiday period using NHITS model. \vspace{2mm}
    
    \item Seasonality decomposition (SD) : As holidays are an annual event, we use MSTL \cite{bandara2021mstl} to decompose the time series and try to extract the yearly seasonality corresponding to each holidays and use the extracted yearly seasonality as estimation of the sales during holiday period.
\end{itemize}
\vspace{-2.5em}

\begin{figure}[H]
    \centering
    \includegraphics[width=1\linewidth]{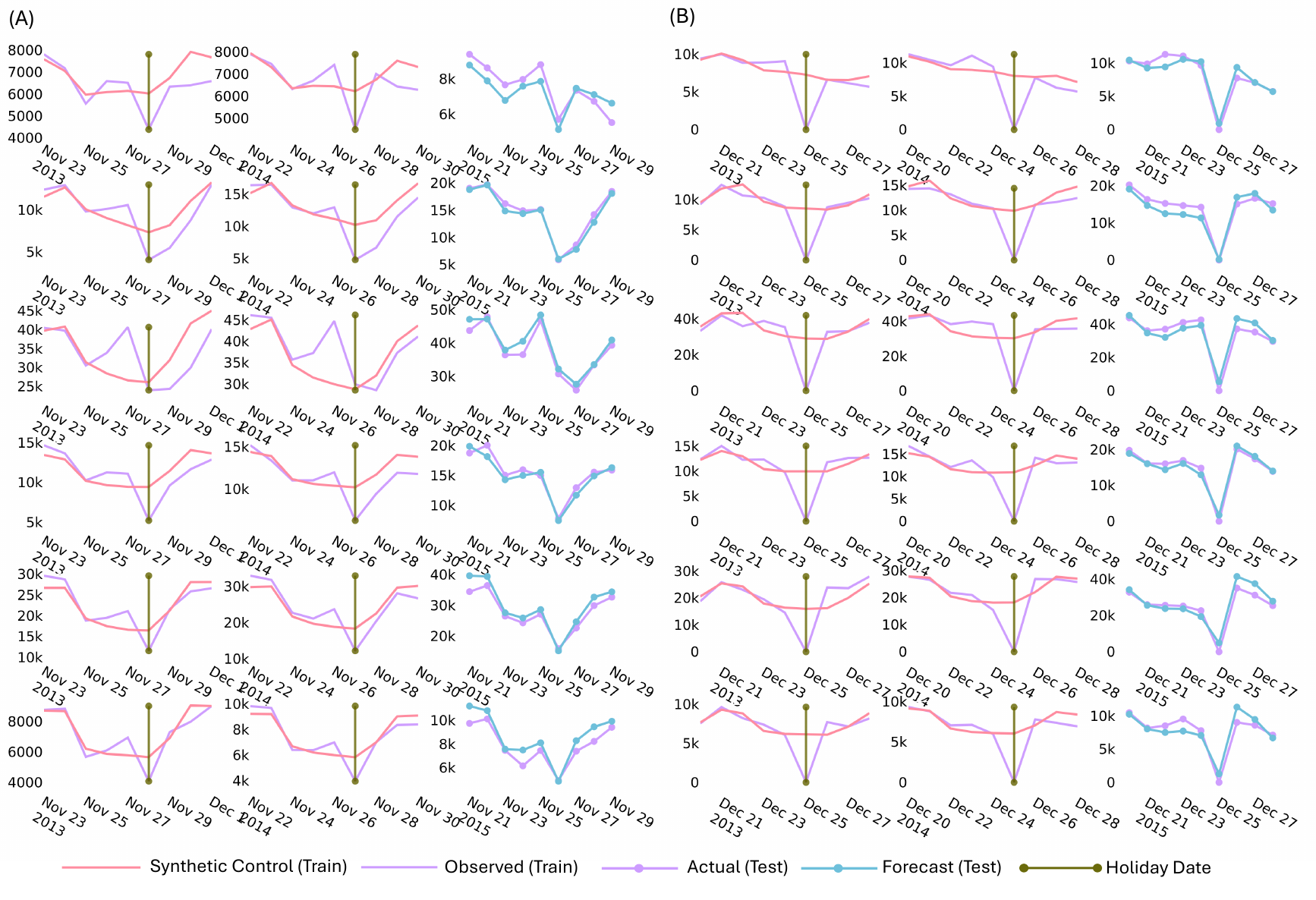}
    \caption{(A). Thanksgiving. (B).Christmas. Each row corresponds to one department.}
    \label{fig:synthetic_control}
\end{figure}

\subsection{Consistency of Estimated Synthetic control}
Figure \ref{fig:synthetic_control} illustrates the experimental results for Thanksgiving (A) and Christmas (B) over multiple years and departments. Each row represents one department. We can see from the first two columns in Fig.\ref{fig:synthetic_control}(A) and Fig.\ref{fig:synthetic_control}(B) that the estimated synthetic control remains consistent across years, demonstrating that our assumption, namely, that the model effectively captures the main signal and is robust to anomalies, holds true in this scenario. The significant deviation between the synthetic control and the observed values is attributed to the holiday impact, which also exhibits a consistent pattern over the years.


\vspace{-1em}
\begin{table}[H]
\centering
\renewcommand{\arraystretch}{1.1}
\scriptsize 
\begin{tabularx}{\textwidth}{c|>{\centering\arraybackslash}X>{\centering\arraybackslash}X>{\centering\arraybackslash}X|>{\centering\arraybackslash}X>{\centering\arraybackslash}X>{\centering\arraybackslash}X}
\hline
\textbf{Department} & \multicolumn{3}{c|}{\textbf{Thanksgiving}} & \multicolumn{3}{c}{\textbf{Christmas}} \\ \hline
& SD & DF & Ours & SD & DF & Ours \\ \hline
Food 1     & 52.60   & 20.21   & \textcolor{red}{8.79}   & 20.24   & 46.17   & \textcolor{red}{7.08}   \\ 
Food 2      & 19.49   & 12.66   & \textcolor{red}{4.24}   & 31.70  & 18.82  & \textcolor{red}{12.77}  \\ 
Food 3      & 18.17   & 21.08   & \textcolor{red}{4.85}   & 12.03   & 24.21   & \textcolor{red}{8.90}   \\ \hline
Hobby 1     & 101.07   & 253.07   & \textcolor{red}{5.67}   & 39.21   & 24.52   & \textcolor{red}{5.23}   \\ 
\hline
Household 1 & 10.21   & 9.33   & \textcolor{red}{7.50}   & 13.11  & 14.28  & \textcolor{red}{10.20}  \\ 
Household 2 & 28.99   & 58.20   & \textcolor{red}{9.74}   & 16.20  & 19.77   & \textcolor{red}{10.75}   \\ \hline
\end{tabularx}
\vspace{1mm}
\caption{Forecast error during holidas. The best performances are mentioned in \textcolor{red}{red}.}
\label{tab:mape_table}
\end{table}
\vspace{-3em}

\subsection{Holiday impact Prediction}
We further evaluate the accuracy of estimated treatment effect using 2015 as test data. To account for the multiplicative property of time series, we first calculate the holiday impact ratio $\mathbf{r}_{i \in \{2012,2013,2014\}}$ by dividing the estimated treatment effect by their year-specific scale $C_{i \in \{2012,2013,2014\}}$, where $C_i$ can be any monthly aggregation to reflect the scale of that year. The average holiday impact ratio $\bar{\mathbf{r}}$ from 2012 to 2014 was then used as the estimated holiday impact ratio for 2015, and $\bar{\mathbf{r}} \times \hat{C}_{2015}$ will be used as the estimated holiday effect for 2015. The figures in the third columns of Fig.\ref{fig:synthetic_control}(A) and Fig.\ref{fig:synthetic_control}(B) demonstrate the accurate forecast for Thanksgiving and Christmas. We also compared MAPE between different methods and it is evident that our method can achieve a significant improvement, as shown in Table \ref{tab:mape_table}. These results align with our intuition that seasonality decomposition-based approaches are sensitive to other confounders in the time series, as many holidays do not exhibit exact periodicity. Out-of-sample forecasts of synthetic control within holiday regions are heavily dependent on the stationarity assumption, which limits their applicability in real-world scenarios.

\section{Conclusion}

In this paper, we address the challenging problem of estimating treatment effects in time series data when the control group is always unobservable during the treatment period, a common scenario such as evaluating the impact of holidays on retail sales. We developed an innovative estimation framework for imputing the always-unobserved control group. We also provided the asymptotic analysis of the treatment effect. Our experimental results demonstrated the robustness and accuracy of our method in capturing impact of events.

\begin{credits}
\subsubsection{\discintname}
There is no conflict of interests.
\end{credits}
%
%
%
\bibliographystyle{splncs04}
\bibliography{ref}

\section*{Appendix}

\subsection*{Proof of Theorem \ref{thm:asymptotic_normality}}

To make the proof go smoothly, we first introduce a necessary lemma that can facilitate this proof.  
\begin{lemma}
\label{lm: consistent}
Let $\hat{\phi}$ be the Ordinary Least Squares (OLS) estimator of $\phi$ based on the observations $\{Y_{i,t}\}_{t=0}^{T_0}$ for $i = 1, \dots, N$. Then, as $N \to \infty$ (with $T_0$ fixed), $\hat{\phi}$ is a consistent estimator of $\phi$, and moreover,
\[
\hat{\phi} - \phi = O_p(N^{-1/2}).
\]
\end{lemma}
Then we start with the proof of Theorem \ref{thm:asymptotic_normality}. 

\begin{proof}
To establish the asymptotic normality of the estimator $\hat{\delta}$, we analyze the estimation error and its distribution as $N \to \infty$.

For each $t = T_0 + 1, \dots, T_0 + d$, the estimation error can be expressed as:
\begin{align*}
\hat{\delta}_t - \delta_t &=  \left( \bar{Y}_t(1) - \hat{\bar{Y}}_t(0) \right) - \delta_t = \left( \bar{Y}_t(0) + \delta_t - \hat{\bar{Y}}_t(0) \right) - \delta_t = \bar{Y}_t(0) - \hat{\bar{Y}}_t(0),
\end{align*}
where $\bar{Y}_t(0) = \dfrac{1}{N} \sum_{i=1}^N Y_{i,t}(0)$ is the sample mean of the potential outcomes without treatment, and $\hat{\bar{Y}}_t(0) = \dfrac{1}{N} \sum_{i=1}^N \hat{Y}_{i,t}(0)$ is the sample mean of the estimated counterfactual outcomes. We then define the individual prediction error as 
\begin{equation*}
    e_{i,t} = Y_{i,t}(0) - \hat{Y}_{i,t}(0), \quad \text{and} \quad \bar{e}_t = \frac{1}{N} \sum_{i=1}^N e_{i,t}.
\end{equation*}
Thus, the causal effect estimation error is simplified to
\begin{equation*}
    \hat{\delta}_t - \delta_t = \bar{e}_t.
\end{equation*}
We then analyze $\bar{e}_t$ and determine its asymptotic distribution. From the AR(1) model and the estimation procedure, the dynamics of the true and estimated counterfactuals are given by: $Y_{i,t}(0) = \phi Y_{i,t-1}(0) + \varepsilon_{i,t}$ and $\hat{Y}_{i,t}(0) = \hat{\phi} \hat{Y}_{i,t-1}(0)$, respectively. Then subtracting the estimated equation from the true equation, we obtain the recursive relationship for the prediction error:
\begin{align*}
e_{i,t} = Y_{i,t}(0) - \hat{Y}_{i,t}(0) & = \phi Y_{i,t-1}(0) - \hat{\phi} \hat{Y}_{i,t-1}(0) + \varepsilon_{i,t} \\
& = \phi e_{i,t-1} - \theta \hat{Y}_{i,t-1}(0) + \varepsilon_{i,t},
\end{align*}
where we denote $\theta = \hat{\phi} - \phi$. At time $t = T_0 + 1$, the initial condition is given by $\hat{Y}_{i,T_0}(0) = Y_{i,T_0}$, since $\hat{Y}_{i,T_0}(0)$ is initialized with the observed value at $T_0$, and $Y_{i,T_0}(0) = Y_{i,T_0}$ because there is no treatment effect before $T_0 + 1$. Therefore, the initial prediction error is $e_{i,T_0} = Y_{i,T_0}(0) - \hat{Y}_{i,T_0}(0) = 0$. At $t = T_0 + 1$, the prediction error becomes
\begin{equation*}
    e_{i,T_0 + 1} = \phi e_{i,T_0} - \theta \hat{Y}_{i,T_0}(0) + \varepsilon_{i,T_0 + 1} = -\theta Y_{i,T_0} + \varepsilon_{i,T_0 + 1}.
\end{equation*}
Since $\theta = \hat{\phi} - \phi = O_p(N^{-1/2})$ according to Lemma \ref{lm: consistent}, and $\hat{Y}_{i,T_0}(0) = Y_{i,T_0}$ is bounded (due to stationarity and finite variance), the term $\theta Y_{i,T_0}$ is of order $O_p(N^{-1/2})$. This term is negligible compared to $\varepsilon_{i,T_0 + 1}$, which is $O_p(1)$. Therefore, we know that
\begin{equation*}
    e_{i,T_0 + 1} \approx \varepsilon_{i,T_0 + 1}.
\end{equation*}
Similarly, for subsequent periods, we can neglect the term involving $\theta$ in the recursive equation for $e_{i,t}$ because it remains of order $O_p(N^{-1/2})$, while the other terms are of order $O_p(1)$. Thus, the prediction error can be approximated by
\begin{equation*}
    e_{i,t} \approx \phi e_{i,t-1} + \varepsilon_{i,t}, \quad t \geq T_0 + 1.
\end{equation*}
Solving this homogeneous linear difference equation with initial condition $e_{i,T_0} = 0$, and we obtain:
\begin{equation*}
    e_{i,t} \approx \sum_{k = 0}^{t - T_0 - 1} \phi^k \varepsilon_{i,t - k},
\end{equation*}
and then we have 
\begin{equation*}
    \bar{e}_t = \frac{1}{N} \sum_{i=1}^N e_{i,t} \approx \sum_{k = 0}^{t - T_0 - 1} \phi^k \bar{\varepsilon}_{t - k},
\end{equation*}
where $\bar{\varepsilon}_{t - k} = \dfrac{1}{N} \sum_{i=1}^N \varepsilon_{i,t - k}$. Since the $\varepsilon_{i,t}$ are i.i.d. with mean zero and variance $\sigma^2$, by the Central Limit Theorem, for each fixed $t$ and $k$, we have
\[
\sqrt{N} \bar{\varepsilon}_{t - k} \xrightarrow{d} \mathcal{N}(0, \sigma^2).
\]
Moreover, the $\bar{\varepsilon}_{t - k}$ are asymptotically independent across different time indices because the $\varepsilon_{i,t}$ are independent over time and across individuals. Therefore, the scaled estimation error is:
\begin{equation*}
    \sqrt{N} (\hat{\delta}_t - \delta_t) = \sqrt{N} \bar{e}_t \approx \sum_{k = 0}^{t - T_0 - 1} \phi^k \sqrt{N} \bar{\varepsilon}_{t - k}.
\end{equation*}
This expression is a linear combination of asymptotically independent normal random variables. Hence, $\sqrt{N} (\hat{\delta}_t - \delta_t)$ is asymptotically normally distributed with mean zero.

Next , we know the variance of $\sqrt{N} (\hat{\delta}_t - \delta_t)$ is given by
\begin{equation*}
    \operatorname{Var}\left( \sqrt{N} (\hat{\delta}_t - \delta_t) \right) = \sigma^2 \sum_{k = 0}^{t - T_0 - 1} \phi^{2k} = \sigma^2 \frac{1 - \phi^{2(t - T_0)}}{1 - \phi^2}.
\end{equation*}
As $t - T_0$ increases, the term $\phi^{2(t - T_0)}$ tends to zero (since $|\phi| < 1$), and the variance approaches $\sigma^2/1 - \phi^2$. Similarly, for $t \neq s$, the covariance between $\sqrt{N} (\hat{\delta}_t - \delta_t)$ and $\sqrt{N} (\hat{\delta}_s - \delta_s)$ is:
\[
\operatorname{Cov}\left( \sqrt{N} (\hat{\delta}_t - \delta_t), \sqrt{N} (\hat{\delta}_s - \delta_s) \right) = \sigma^2 \sum_{k = 0}^{t - T_0 - 1} \sum_{l = 0}^{s - T_0 - 1} \phi^{k + l} \operatorname{Cov}\left( \bar{\varepsilon}_{t - k}, \bar{\varepsilon}_{s - l} \right).
\]
Since $\bar{\varepsilon}_{t - k}$ and $\bar{\varepsilon}_{s - l}$ are uncorrelated unless $t - k = s - l$, the covariance reduces to:
\[
\operatorname{Cov}\left( \sqrt{N} (\hat{\delta}_t - \delta_t), \sqrt{N} (\hat{\delta}_s - \delta_s) \right) = \sigma^2 \sum_{k = 0}^{t - T_0 - 1} \phi^{2k + t - s} \delta_{t - k, s - l},
\]
where $\delta_{t - k, s - l}$ is the Kronecker delta function. Note that this covariance is non-zero only when $t - k = s - l$, which implies $t - s = k - l$. However, since $k$ and $l$ range from $0$ to finite numbers (less than $d$), and $t \neq s$, the number of terms where $t - k = s - l$ is limited and does not grow with $N$. Therefore, the covariance between $\sqrt{N} (\hat{\delta}_t - \delta_t)$ and $\sqrt{N} (\hat{\delta}_s - \delta_s)$ tends to zero as $N \to \infty$. Thus, the covariance matrix $\Sigma$ is asymptotically diagonal with diagonal elements:
\[
\Sigma_{t,t} = \operatorname{Var}\left( \sqrt{N} (\hat{\delta}_t - \delta_t) \right) = \frac{\sigma^2}{1 - \phi^2}.
\]
In conclusion, the vector $\sqrt{N} (\hat{\delta} - \delta)$ converges in distribution to a multivariate normal distribution
\[
\sqrt{N} (\hat{\delta} - \delta) \xrightarrow{d} \mathcal{N} \left( \mathbf{0}, \Sigma \right),
\]
where $\Sigma$ is a diagonal matrix with elements $\Sigma_{t,t} = \dfrac{\sigma^2}{1 - \phi^2}$, and we thus complete the proof of Theorem \ref{thm:asymptotic_normality}.
\end{proof}

\subsection*{Proof of Lemma \ref{lm: consistent}}

\begin{proof}
We aim to show that the OLS estimator $\hat{\phi}$ satisfies $\hat{\phi} - \phi = O_p(N^{-1/2})$ as $N \to \infty$, with $T_0$ fixed. We first recall that the OLS estimator $\hat{\phi}$ is given by
\begin{equation*}
    \hat{\phi} = \frac{\sum_{i=1}^N \sum_{t=1}^{T_0} Y_{i,t-1} Y_{i,t}}{\sum_{i=1}^N \sum_{t=1}^{T_0} Y_{i,t-1}^2}.
\end{equation*}
We can rewrite the numerator using the model equation:
\begin{align*}
\sum_{i=1}^N \sum_{t=1}^{T_0} Y_{i,t-1} Y_{i,t} =\phi \sum_{i=1}^N \sum_{t=1}^{T_0} Y_{i,t-1}^2 + \sum_{i=1}^N \sum_{t=1}^{T_0} Y_{i,t-1} \varepsilon_{i,t}.
\end{align*}
Therefore, the estimator can be expressed as:
\begin{equation*}
    \hat{\phi} = \phi + \underbrace{\frac{\frac{1}{N}\sum_{i=1}^N \sum_{t=1}^{T_0} Y_{i,t-1} \varepsilon_{i,t}}{\frac{1}{N}\sum_{i=1}^N \sum_{t=1}^{T_0} Y_{i,t-1}^2}}_{I}.
\end{equation*}
By the Law of Large Numbers (weak), we have 
$$\frac{1}{N}\sum_{i=1}^N \sum_{t=1}^{T_0} Y_{i,t-1}^2 \xrightarrow{p} T_0 E[Y_{i,t-1}^2],$$ 
as $N \to \infty$, where $E[Y_{i,t-1}^2]$ is finite and positive, with $T_0$ constant, we have the denomitor of $I$ is of order $O(1)$. On the other hand, we have the numeritor of $I$ is a sum of mean-zero random variables and the variance is given by 
\begin{align*}
\frac{1}{N^2}\sum_{i=1}^N \operatorname{Var}\left( \sum_{t=1}^{T_0} Y_{i,t-1} \varepsilon_{i,t} \right) = \frac{1}{N^2}\sum_{i=1}^N \sum_{t=1}^{T_0} \operatorname{Var}\left( Y_{i,t-1} \varepsilon_{i,t} \right) = \frac{\sigma^2}{N^2} \sum_{i=1}^N \sum_{t=1}^{T_0} E[Y_{i,t-1}^2].
\end{align*}
Then we know that by the Chebychev's inequality, we have the $I = O_p(1/\sqrt{N})$, and thus we finish the proof of Lemma \ref{lm: consistent}
\end{proof}

\end{document}